\documentclass[
 aip,
 apl,
 amsmath,
 amssymb,
 reprint,
]{revtex4-2}

\usepackage{graphicx}
\usepackage{dcolumn}
\usepackage{bm}

\usepackage{etoolbox}

\usepackage[utf8]{inputenc}
\usepackage[T1]{fontenc}
\usepackage[colorlinks=true, citecolor=blue]{hyperref}
\usepackage{units}
\usepackage[dvipsnames]{xcolor}

\newcommand{\angstrom}{\text{\AA}\vphantom{A}}
\newcommand{\kB}{k_\mathrm B}
\newcommand{\Tf}{T_\mathrm{Pb}}
\newcommand{\Ts}{T_\mathrm{Si}}
\newcommand{\DT}{\Delta T}
\newcommand{\tc}{\tau_\mathrm{cool}}

\newcommand{\trs}{\tau_\mathrm{rec}}

\newcommand{\Dt}{\Delta t}
\newcommand{\DI}{\Delta I}

\usepackage{changes}
\definechangesauthor[name=HvH, color=Peach]{HvH}
\definechangesauthor[name=CB, color=WildStrawberry]{CB}

\graphicspath{ {./images/} }

\makeatletter

\def\@email#1#2{%
 \endgroup
 \patchcmd{\titleblock@produce}
  {\frontmatter@RRAPformat}
  {\frontmatter@RRAPformat{\produce@RRAP{*#1\href{mailto:#2}{#2}}}\frontmatter@RRAPformat}
  {}{}
}%

\makeatother

\begin{document}

\preprint{AIP/123-QED}

\title[]{Thermal boundary conductance under large temperature discontinuities of ultrathin epitaxial Pb films on Si(111)}


\author{Christian Brand}
\email{christian.brand@uni-due.de}
\affiliation{Faculty of Physics, University of Duisburg-Essen, 47057 Duisburg, Germany}
\affiliation{Institut für Festkörperphysik, Leibniz Universität Hannover, 30167 Hannover, Germany}

\author{Tobias Witte}
\affiliation{Faculty of Physics, University of Duisburg-Essen, 47057 Duisburg, Germany}

\author{Mohammad Tajik}
\affiliation{Faculty of Physics, University of Duisburg-Essen, 47057 Duisburg, Germany}

\author{\firstname{Jonas D.} Fortmann}
\affiliation{Faculty of Physics, University of Duisburg-Essen, 47057 Duisburg, Germany}

\author{Birk Finke}
\affiliation{Faculty of Physics, University of Duisburg-Essen, 47057 Duisburg, Germany}

\author{Herbert Pfnür}
\affiliation{Institut für Festkörperphysik, Leibniz Universität Hannover, 30167 Hannover, Germany}

\author{Christoph Tegenkamp}
\affiliation{Institut für Festkörperphysik, Leibniz Universität Hannover, 30167 Hannover, Germany}
\affiliation{Institut für Physik, Technische Universität Chemnitz, 09126 Chemnitz, Germany}

\author{Michael \surname{Horn-von Hoegen}}
\affiliation{Faculty of Physics, University of Duisburg-Essen, 47057 Duisburg, Germany}
\affiliation{Center for Nanointegration (CENIDE), University of Duisburg-Essen, 47057 Duisburg, Germany}

\date{\today}


\begin{abstract}
Heat transfer is a critical aspect of modern electronics, and a deeper understanding of the underlying physics is essential for building faster, smaller, and more powerful devices with an improved performance and efficiency.
In such nanoscale structures, the heat transfer between two materials is limited by the finite thermal boundary conductance across their interface.
Using ultrafast electron diffraction under grazing incidence we investigated the heat transfer from ultrathin epitaxial Pb films to an Si(111) substrate under strong non-equilibrium conditions.
Applying an intense femtosecond laser pulse, the \unit[5-7]{ML} thin Pb film experiences a strong heat up by \unit[10-120]{K} while the Si substrate remains cold at $\approx \unit[10]{K}$.
At such large temperature discontinuities we observe a significantly faster cooling for stronger excited Pb films.
The decrease of the corresponding cooling time constant is explained through the thermal boundary conductance in the framework of the diffuse mismatch model.
The thermal boundary conductance is reduced by more than a factor of three in comparison with Pb films grown on H-terminated substrates, pointing out the importance of the morphology of substrate, heterofilm and their interface.
\end{abstract}

\maketitle





\par
With down-scaling the spatial dimensions in modern electronic devices into the nanometer regime, heat dissipation from an electronically active medium across an interface towards the substrate is no longer governed by the thermal conductivity $\kappa$ of the bulk but rather limited by the thermal boundary conductance $G$ of the interface \cite{lyeo2006thermal, cahill2003nanoscale, cahill2014nanoscale}.
The discontinuity of the elastic properties, i.e., the sound velocities and the phonon density of states (PDOS) of film and substrate lead to an additional resistance in heat transport, which is accompanied by a discontinuity in temperature $\DT$ at the interface \cite{swartz1989thermal}.
For films thinner than the Kapitza length $\kappa / G$ ($\approx \unit[10]{\mu m}$ for Pb/Si at \unit[100]{K}), the heat transfer is governed by the interfacial conductance, whereas for thicker films the bulk thermal conductivity dominates.

\par 
Usually $G$ is determined by techniques such as time domain \cite{cahill1990thermal, cahill2004analysis, lyeo2006thermal, schmidt2013pump, Schmidt:RSI.79.114902, Chang:APL125.212201} and frequency domain \cite{Malen:JHT133.081601, Regner:RSI84.064901, Yang:RSI84.104904} thermoreflectance or $3 \omega$ method \cite{Cahill:PRB35.4067, cahill1990thermal, Dames:ARHT16.7} under conditions where the equilibrium between film and substrate is only slightly distorted.
Thus, the temperature difference $\DT$ between film and substrate is small compared to the temperature of the substrate.
However, such a situation might be untypical for technological applications, e.g., electronic devices in which heat dissipation is strongly localized with spatial dimensions in the nanometer regime, and large temperature discontinuities may occur at the interfaces with heat fluxes up to $\unit[100]{kW/cm^2}$ \cite{Hu:JoAP104.083503}.
Under such conditions, the heat transfer across interfaces is no longer defined by a constant $G$ but becomes temperature-dependent.
Cases with a large temperature discontinuity at the interface, however, are not accessible under commonly used experimental setups such as the thermoreflectance or the $3 \omega$ method.

\par
\begin{figure*}[ht]
\centering
\includegraphics[width=1.0\textwidth]{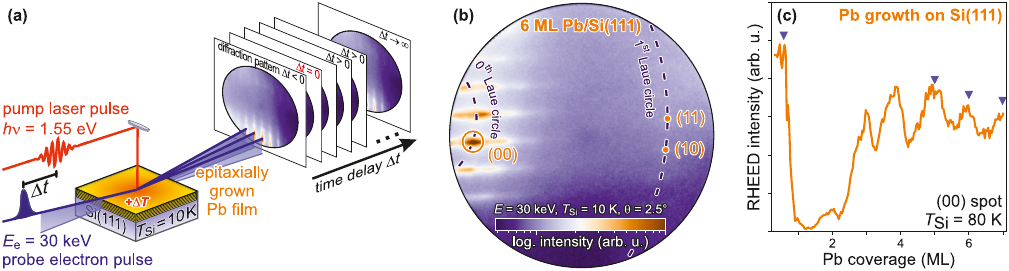}
\caption{\textbf{URHEED experiment:}
(a) Scheme of the laser pump - electron probe setup for the Pb/Si(111) heterosystem.
(b) RHEED pattern at an electron energy of \unit[30]{keV} taken from a \unit[6]{ML} thin epitaxial Pb(111) film grown on Si(111).
(c) Thickness calibration through layer-by-layer RHEED intensity oscillations of the (00) spot as function of Pb coverage during growth by MBE.
The blue triangles indicate the positions where the shutter of the evaporator was opened and closed, i.e., at \unit[1/3]{ML} Pb coverage (Si(111)-$\beta(\sqrt{3} {\times} \sqrt{3})\mathrm{R}30^\circ$-Pb wetting layer), and at 5, 6, and \unit[7]{ML} Pb coverage, respectively. 
}
\label{Fig.Scheme}
\end{figure*}

 
Here, we used ultrafast reflection high energy electron diffraction (URHEED) as thermometer to follow the transient temperature evolution of the heterofilm upon impulsive excitation through a femtosecond laser pulse \cite{krenzer2006thermal, hanisch2008thermal, hanisch2013ultra, frigge2015nanoscale, witte2017nanoscale, Gorfien:StructDyn7.025101} [cf.\ Fig.~\ref{Fig.Scheme}~(a)].
RHEED at grazing incidence ensured surface sensitivity \cite{braun1999applied} and large intensity changes upon changes of temperature on the order of $I/I_0 \approx 0.5\%$ per Kelvin.
We employed the Debye-Waller effect to determine the transient change of temperature $\DT(\Dt)$ from the drop of diffraction spot intensity $\DI(\Dt)$ \cite{debye1913interferenz, waller1923frage}.

\par
Here, we nicely turned a seemingly disadvantage into an advantage: the URHEED technique requires high excitation density, i.e., large temperature rises $\DT$ on the order of \unit[10-120]{K}, subsequent to a short laser pulse for a sufficient signal-to-noise ratio, which is ultimately limited by the point-to-point stability and longtime drift of the fs-laser pulses from the laser amplifier.
This allowed us to determine the thermal boundary conductance under extreme non-equilibrium conditions at large temperature discontinuities $\DT \gg \Ts$, i.e., at a Si substrate temperature of $\Ts = \unit[10]{K}$ the temperature rise of the film was more than ten times higher than the substrate's temperature in case of the highest incident laser pulse energy used!

\par
All experiments were performed under ultra-high vacuum (UHV) conditions at a base pressure of \unit[$4 {\times} 10^{-10}$]{mbar}.
The Si(111) substrate ($\pm 0.1^\circ$, $\unit[380]{\mu m}$, \unit[0.6-1]{$\Omega$\,cm}, $n$-doped with \unit[6-10${\times} 10^{15}$]{P atoms per cm$^3$}, Virginia Semiconductor) was mounted on a cryostat cooled by liquid He.
The substrate was prepared by degassing at \unit[600]{$^\circ$C} for several hours followed by flash-annealing at \unit[1250]{$^\circ$C} to desorb the native oxide and prepare a clean $(7 {\times} 7)$-reconstructed surface.
The Pb film was grown by molecular beam epitaxy (MBE) from an indirectly heated Knudsen-type ceramic crucible \cite{kury2005compact}.
As a template for film deposition, a \linebreak \mbox{Si(111)-$\beta(\sqrt{3} {\times} \sqrt{3})\mathrm{R}30^\circ$-Pb} reconstruction 
was prepared through Pb deposition and followed by annealing for \unit[20]{s}, both at $\unit[600]{^\circ\mathrm{C}}$.
Subsequently, using the kinetic pathway, continuous epitaxial Pb films of \mbox{$d$ = \unit[5-7]{ML}} thickness (1\,monolayer (ML) Pb(111) $d = \unit[2.86]{\angstrom}$, \mbox{$\unit[1.02 \times 10^{15}]{atoms/cm^2}$}) were grown at low sample temperature $< \unit[80]{K}$ [cf.\ RHEED pattern in Fig.~\ref{Fig.Scheme}~(b)], thus avoiding islanding of the Pb films \cite{petkova2001order}, and resulting in a smooth surface and high quality of the Pb films as proven by {\it in-situ} low energy electron diffraction.
The film thickness was calibrated through RHEED intensity oscillations of the (00) spot during layer-by-layer growth \cite{Ichimiya_Cohen_2004, horn1999growth} as shown in Fig.~\ref{Fig.Scheme}~(c).

\begin{figure*}[t!]
\centering
\includegraphics[width=1.0\textwidth]{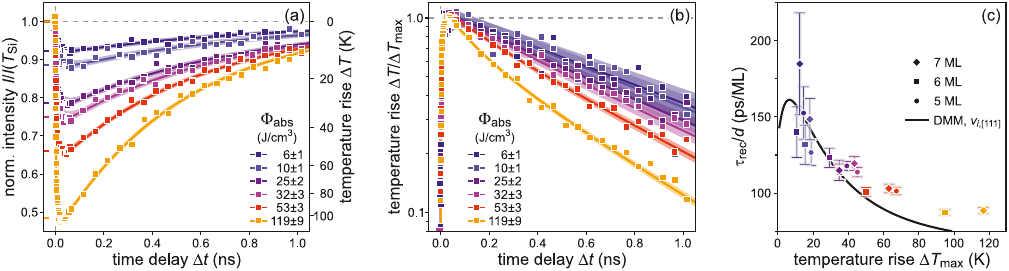}
\caption{\textbf{Lattice dynamics of ultrathin Pb films:}
(a) Excitation and recovery of the normalized intensity $I(\Dt)/I(\Ts)$ of a \unit[6]{ML} thin film using different absorbed energy densities $\Phi_\mathrm{abs}$.
(b) Corresponding transient normalized temperature change $\DT(\Dt)/\DT_\mathrm{max}$.
(c) Fraction of recovery time constant $\trs$ and film thickness $d$ as function of temperature rise $\DT_\mathrm{max}$ from experiment (data points) and from DMM (black curve).
The color of the symbols indicates the excitation strength, with blue corresponding to weak and orange to strong excitation.
The shaded areas in (a,b) are the $1\sigma$ uncertainties of the fits.
}
\label{Fig.results}
\end{figure*}

\par
The ultrathin Pb films were impulsively excited by \unit[80]{fs} laser pulses at a photon energy of \unit[1.55]{eV} with a repetition rate of \unit[5]{kHz} at normal incidence [Fig.~\ref{Fig.Scheme}~(a)].
Photons at this energy excite the metallic Pb film to much higher excess energy density than the substrate because for Si the large direct band gap of \unit[3.4]{eV} and the absorption length \cite{Aspnes:PhysRevB27.985} of $\unit[12.7]{\mu m}$ suppress an effective optical excitation of the substrate.
The variation of the laser pulse energy by a factor of $\approx 25$ (incident fluence \mbox{$F_\mathrm{in}$ = \unit[0.05-1.36]{W/cm$^2$}}) resulted in a variation of the absorbed energy density $\Phi_\mathrm{abs}$ in the Pb films from 6 to $\unit[140]{J/cm^3}$.
Since the diameter of the pump laser beam (\unit[6]{mm}) was much larger than the width of the sample (\unit[2]{mm}) a homogeneous excitation of the film was ensured.
The sample base temperature of the instrument was measured by a Si diode (Lake Shore Cryotronics \mbox{DT-670C-SD}) soldered by In to a different Si sample with the same size and specifications as used during the experiments.
Upon liquid He cooling we determined a minimum temperature of \unit[19]{K} under high vacuum conditions for this sample.
During all experiments the Cu mount connected to the cold head of the cryostat was as cold as \unit[9]{K} as determined by another Si diode (Lake Shore Cryotronics \mbox{DT-670B-CU-HT}).
Due to heating by radiation and by the wires to the Si diodes we expect the real sample base temperature below \unit[19]{K} and above \unit[9]{K}.
Based on these constraints and the results of our data analysis, we estimated a sample base temperature of $\Ts \approx \unit[10^{+5}_{-1}]{K}$.
The difference in heat between \unit[9]{K} and \unit[19]{K} for the electron and lattice system amounts to $\unit[3.8]{J/cm^3}$ which is far below the applied absorbed energy densities in the experiment.
The high thermal conductivity of up to $\approx \unit[3500]{W/m\,K}$ \cite{Thompson:JPhysChemSol20.146, glassbrenner1964thermal} of slightly $n$-doped Si at $\Ts$ renders the substrate as an almost perfect sink for the dissipated heat from the Pb film, i.e., the substrate's temperature rise was on the order of \unit[1-2]{K} only \cite{hanisch2021violation}.
Further details on the experimental setup can be found elsewhere \cite{krenzer2006thermal, hafke2019pulsed, HvH:StrctDyn11.021301}.

\par
The Debye-Waller effect
$I = I_0 \exp \left( -\frac{1}{3} \langle {\bf u}^2 \rangle \cdot {\bf \Delta k}^2 \right)$
for the intensity $I$ of a diffraction spot with isotropic mean squared vibrational displacements $\langle {\bf u}^2 \rangle$ of the atoms and momentum transfer ${\bf \Delta k}$ is employed in URHEED at an electron energy of \unit[30]{keV} to follow the transient temperature $\Tf$ of the Pb films upon optical excitation in a pump-probe setup as sketched in Fig.~\ref{Fig.Scheme}~(a).
We used a grazing angle of incidence of $\theta = 2.5^\circ {\pm} 0.1^\circ$ to provide surface sensitivity, i.e., $\left| {\bf \Delta k} \right| = \unit[7.6 {\pm} 0.3]{\angstrom^{-1}}$ for the (00) spot.
$\Tf$ is determined from $\langle {\bf u}^2 \rangle$ in the framework of the Debye model.
In order to account for the strong thermal lattice expansion of Pb \cite{Rubin:JPC66.266} we used the temperature-dependent equivalent Debye temperature $\Theta_\mathrm{Pb,b}$ of bulk Pb \cite{Gupta:PhysicaBC122.236} for the analysis.
We also took into account that in the URHEED experiment mostly the topmost atomic layer of the Pb films was probed which exhibits a reduced surface Debye temperature $\Theta_\mathrm{Pb,s} \approx 0.68 \Theta_\mathrm{Pb,b}$ \cite{elsayed1996surface, Tinnemann:StructDyn6.2329} while the underlying layers were considered as bulk.
Thus, the (00) spot intensity at temperature $\Tf$ is described by
\begin{equation}
\label{eq:Intensity}
    \frac{I(\Tf)}{I_0} = \exp \left[ - \gamma \left(1 + \frac{4 \Tf^2}{\Theta_\mathrm{Pb,s}^2} \int_0^{\frac{\Theta_\mathrm{Pb,s}}{\Tf}} \frac{x\,\mathrm{d}x}{e^x - 1} \right) \right]\,,
\end{equation}
where $\gamma = 3 \hbar^2 {\bf \Delta k}^2/4 m_\mathrm{Pb} \kB \Theta_\mathrm{Pb,s}$ and $m_\mathrm{Pb}$ is the mass of the Pb atoms.

\par
The lattice excitation occurs subsequently to fs-optical irradiation in \unit[3]{ps} through electron-phonon coupling of the photo-excited carriers followed by equilibration of the phonons through anharmonic coupling in \unit[20-90]{ps} \cite{Tajik:arXiv2312.04541}.
While the details of the excitation process are beyond the scope of this Letter, the thermal boundary conductance $G$ is determined by the cooling behavior of the film.
Thus, the evolution of film temperature $\Tf$ is evaluated through the transient intensity at time delay $\Dt$ between laser pump and electron probe pulse by fitting an exponential function for the recovery as
\begin{equation}
\label{Eq:DeltaI}
    I(\Dt)/I(\Ts) = \delta I f_\mathrm{exc}(\Dt) e^{-\Dt/\trs}\,,
\end{equation}
where $\delta I$ is a fitting parameter for the maximum intensity drop, $f_\mathrm{exc}$ describes the excitation process of the lattice, and $\trs$ is the recovery time constant, respectively.

\par
As the substrate temperature remains almost constant, $G$ can be determined by \cite{stoner1993kapitza, krenzer2006thermal, hanisch2021violation} 
\begin{equation}
    G = - \frac{c_\mathrm{Pb} \varrho_\mathrm{Pb} d}{\Tf - \Ts} \frac{\partial \Tf}{\partial \Dt}\,,
\label{eq:TBCexp_time}
\end{equation}
where $c_\mathrm{Pb}$ is the specific heat capacity, $\varrho_\mathrm{Pb}$ is the mass density and $d$ is the thickness of the Pb film.
In the high-temperature limit $\Tf \gg \Ts$, the solution of the differential equation for $\Tf$ results in an exponential decrease of the film temperature with a time constant $\tc \approx \trs$.
Under these conditions, Eq.~\eqref{eq:TBCexp_time} simplifies to
\begin{equation}
    G = \frac{c_\mathrm{Pb} \varrho_\mathrm{Pb} d}{\tc}\,.
\label{eq:TBCexp}
\end{equation}
Of course, due to the large temperature rise $\DT$ in our experiment, $c_\mathrm{Pb}$, $\varrho_\mathrm{Pb}$ and $d$ are not constant during cooling of the film from $\Tf$ to $\Ts$ \cite{meads1941heat, Rubin:JPC66.266}.
However, for simplicity of the analysis we used the values at maximum film temperature $\Ts + \DT_\mathrm{max}$.
Furthermore, electronic heat conduction and electron–phonon scattering at the interface are negligible.

\par
\begin{figure}[t]
\centering
\includegraphics[width=1.0\columnwidth]{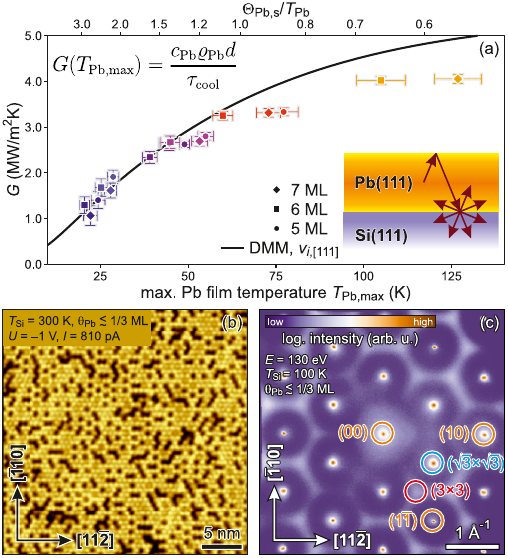}
\caption{\textbf{Thermal boundary conductance $G$ and structure of the wetting layer:}
(a) Experimental $G$ for film thicknesses of \unit[5-7]{ML} and compared to DMM in Debye model with the sound velocities along the [111] direction (black curve).
The color of the symbols indicates the excitation strength, with blue corresponding to weak and orange to strong excitation.
Inset: Scattering mechanism for phonons in DMM.
(b) STM image of the Si(111)-$\beta(\sqrt{3} {\times} \sqrt{3})\mathrm{R}30^\circ$-Pb wetting layer at room temperature taken at negative bias of \unit[-1]{V} with a current of \unit[810]{pA}, i.e., probing occupied states.
Bright features correspond to Pb atoms while dark features indicate Si atoms in substitutional sites replacing the Pb atoms.
The Pb coverage is reduced to \unit[0.28]{ML} instead of \unit[1/3]{ML}.
(c) SPA-LEED pattern of the Si(111)-$\beta(\sqrt{3} {\times} \sqrt{3})\mathrm{R}30^\circ$-Pb wetting layer at \unit[100]{K} taken at \unit[130]{eV}.
The formation of the Pb/Si surface alloy and increased disorder causes the honeycomb-shaped diffuse intensity at $(3 {\times} 3)$ positions.}
\label{Fig.TBC}
\end{figure}

For comparison with theory, we use the semiclassical diffuse mismatch model (DMM) \cite{swartz1989thermal}.
In DMM, phonons are considered to diffusely scatter at the interface with conservation of energy but without conservation of momentum and direction, and thus with arbitrary final momentum and direction (see inset of Fig.~\ref{Fig.TBC}). 
The transmission probability $\Gamma_\mathrm{DMM}$ for phonons to cross the heterointerface is then given by Fermis golden rule: the ratio of PDOS in film and substrate determines the final state in diffuse scattering, i.e., transmission towards the substrate or backscattering into the film.
Taking into account the absence of optical phonon branches in Pb, the phonon transmission probability from the Pb film into the Si substrate is given in the Debye model (where $\mathrm{PDOS}_i \propto v_i^{-3}$) by
\begin{equation}
    \Gamma_\mathrm{DMM} = \frac{1}{2} \left( 1 + 2 \frac{v_{\mathrm{Pb},l}^{-2} + 2v_{\mathrm{Pb},t}^{-2}}{v_{\mathrm{Si},l}^{-2} + 2v_{\mathrm{Si},t}^{-2}} \right)^{-1}\,,
\label{eq:GammaDMM2}
\end{equation}
where the $v_i$ are the longitudinal ($l$) and transversal ($t$) sound velocities of the Pb film and the Si substrate along the [111] direction.
We used bulk values for the phonon density of states and heat capacity as Monte Carlo simulations have shown that even heterofilms consisting of only four ML almost exhibit bulk properties \cite{Krenzer:PhysRevB80.024307}.

\par
Figure~\ref{Fig.Scheme}~(b) depicts a RHEED pattern of a \unit[6]{ML} Pb film.
The transient normalized intensity $I(\Dt)/I(\Ts)$ of the (00) spot is shown in Fig.~\ref{Fig.results}~(a) for different absorbed energy densities $\Phi_\mathrm{abs}$, while a video of $\Delta I(\Dt)/I(\Ts)$ for the whole diffraction pattern is shown in the supplementary material.
Upon laser excitation at time delay $\Dt = 0$, the intensity decreases on a timescale of \unit[26-67]{ps} (time delay of minimum intensity) depending on the excitation strength.
The recovery of the intensity to $I(\Ts)$ (dashed grey line) occurs on a much slower timescale of several hundred picoseconds.
The experimental data were fitted using Eq.~\eqref{Eq:DeltaI} and the fits are shown as solid lines in Fig.~\ref{Fig.results}~(a).
The observed intensity drop becomes larger with increasing $\Phi_\mathrm{abs}$, i.e., the film temperature $\Tf$ is rising higher.
The maximum rise of film temperature $\DT_\mathrm{max}$ as determined from the maximum intensity drop $\DI_\mathrm{max}/I(\Ts)$  is indicated by short colored lines on the left ordinate in Fig.~\ref{Fig.results}~(a) and is ranging from \unit[10.6]{K} for the lowest to \unit[94.8]{K} for the highest absorbed energy density.
In Fig.~\ref{Fig.results}~(b), the transient intensity $I(\Dt)$ is converted into the normalized temperature change $\DT(\Dt)/\DT_\mathrm{max}$.
The fit of $I(\Dt)/I(\Ts)$ yields the time constants $\trs$ for the recovery of the transient intensity.
As shown in Fig.~\ref{Fig.results}~(c), when $\trs$ is normalized by the film thickness $d$, the data collapse onto a single curve which is well described by the DMM (black line).
The cooling significantly speeds up with increasing $\Phi_\mathrm{abs}$, i.e., $\trs$ ranges from $\unit[840 {\pm} 99]{ps}$ for the lowest to $\unit[524 {\pm} 11]{ps}$ for the highest value of $\Phi_\mathrm{abs}$ for $d = \unit[6]{ML}$.

\par
We determined $G$ by employing Eq.~\eqref{eq:TBCexp} as shown in Fig.~\ref{Fig.TBC} for film thicknesses of \unit[5-7]{ML}.
The data for all three thicknesses collapse onto a single curve, which exhibits a clear dependence on the maximum film temperature $T_\mathrm{Pb,max}$, i.e., on the absorbed energy density, with a steeper rise at lower $\DT$ and a significantly slower rise at higher $\DT$.
For $d = \unit[6]{ML}$ the values range from $G =\unit[1.29 {\pm} 0.18]{MW/m^2\,K}$ for the lowest to $\unit[4.02 {\pm} 0.09]{MW/m^2\,K}$ for the highest film temperature rise.
The experimental values nicely match the theoretical expectation for the DMM for temperature increases up to $\DT \approx \unit[50]{K}$.
The slight deviation between experiment and DMM at higher $T_\mathrm{Pb,max}$ might arise from the simplicity of the applied model.
More complex descriptions \cite{Chung:JHT126.376, Hopkins:JHT133.062401, Reddy:APL87.211908, Duda:JApplPhys108.073515} considering the real PDOS and dispersion of the phonons, as well as anharmonic contributions beyond the ones captured by the temperature dependencies of $c_\mathrm{Pb}$, $\Theta_\mathrm{Pb,b}$, $\varrho_\mathrm{Pb}$, $d$ and $v_i$, and also inelastic phonon scattering might further improve the agreement.

\par
Applying the DMM is justified because preparation of the Si(111)-$\beta(\sqrt{3} {\times} \sqrt{3})\mathrm{R}30^\circ$-Pb wetting layer prior to film growth results in a surface alloy.
During preparation at \unit[600]{$^\circ\mathrm{C}$}, part of the Pb atoms desorbed, reducing the coverage from the ideal \unit[1/3]{ML} to only \unit[0.28]{ML}, as evident from the Si atoms in substitutional sites seen in the scanning tunneling microscopy (STM) image in Fig.~\ref{Fig.TBC}~(b).
The accompanying loss of long-range order and increased disorder are reflected in the spot profile analyzing - low energy electron diffraction (SPA-LEED) pattern in Fig.~\ref{Fig.TBC}~(c), showing a diffuse honeycomb intensity at $(3 {\times} 3)$ positions.
This surface alloy of heavy Pb and light Si atoms induces diffuse scattering and relaxes momentum conservation of phonons at the interface and consequently makes DMM mandatory.

\par
The role of crystallinity, grain size, disorder, and defects in the Pb film, as well as the sharpness and well-defined nature of the interface on the thermal boundary conductance $G$ could be assessed by comparison with Pb films grown under different experimental conditions.
Lyeo and Cahill \cite{lyeo2006thermal} studied Pb films grown at room temperature on a H-terminated Si(111) substrate after HF treatment.
Such a preparation results in strongly textured Pb(111) films due to a much rougher Si surface \cite{Angermann:ASS.254.3615, Henrion:ASS202.199}.
H termination results in a significantly more complex interface with modified properties, which leads to a higher phonon transmission probability.
The resulting value for $G$ by Lyeo and Cahill \cite{lyeo2006thermal} of $\approx \unit[14]{MW/m^2\,K}$ at \unit[90-120]{K} is more than three times higher than in our measurements on monocrystalline epitaxial Pb films with a well-defined alloyed but sharp interface.
Thus, high-quality substrate and film morphologies are crucial for lowering the thermal boundary conductance of the Pb/Si interface to the ultimate limit.

\par
In summary, the transient heating of ultrathin epitaxial Pb films on a Si(111) substrate was studied utilizing fs-laser pulses for pumping and ps electron pulses for probing.
The temperature evolution was analyzed applying the Debye-Waller effect to the measured decline in URHEED intensity.
From the data we derive the thermal boundary conductance $G$ of the interface at the maximum film temperature, specifically at large temperature discontinuities up to $\DT \approx \unit[120]{K}$.
The temperature dependence of the thermal boundary conductance matches the result of a simple diffuse mismatch model for low excitation densities.
The thermal boundary conductance of our high-quality epitaxial Pb films grown on an atomically perfectly ordered Si substrate is much lower than for films grown at room temperature on a rough Si surface indicating the crucial role of substrate, film and interface morphologies on the heat transfer.


\section*{Supplementary Material}

\par
See the supplementary material for a video on the transient intensity change in URHEED for a \unit[6]{ML} Pb film upon the highest absorbed energy density.


\section*{Author Contributions}

\par
T.W. performed URHEED and SPA-LEED experiments.
C.B. performed STM and SPA-LEED experiments.
C.B., T.W., and M.T. analyzed the data.
C.B. prepared the figures.
C.B., M.T., J.D.F., B.F., and M.H.-v.H. drafted the manuscript.
M.H.-v.H., H.P. and C.T. conceived and supervised the project.
The manuscript was written through contributions of all authors.
All authors have given approval to the final version of the manuscript.

\par
The authors declare no competing financial interest.


\section*{Acknowledgements}

\par
Funded by the Deutsche Forschungsgemeinschaft (DFG, German Research Foundation) through projects B04 and C03 of the Collaborative Research Center SFB1242 "Nonequilibrium dynamics of condensed matter in the time domain" (Project-ID 278162697) and the project "Transport and collective excitations in metallic wires" of the Research Unit FOR1700 "Metallic nanowires on the atomic scale: Electronic and vibrational coupling in real world systems" (Project-ID 194370842).
The authors thank D.\ G.\ Cahill, B.\ Rethfeld, and Th.\ Groven for fruitful discussions.


\section*{Data Availability}

\par
The data that support the findings of this study are available from the corresponding author upon reasonable request.


\bibliography{Bibliography.bib}


\end{document}